# MESURE DE L'INCERTITUDE TENDANCIELLE SUR LA MORTALITE

## APPLICATION A UN REGIME DE RENTES EN COURS DE SERVICE

**Frédéric PLANCHET*** **Marc JUILLARD**

***ISFA – Laboratoire SAF***
*Université Claude Bernard – Lyon 1*
*50 avenue Tony Garnier*
*69007 LYON*
*FRANCE*

***WINTER & Associés***
*18, avenue Félix Faure*
*69007 Lyon*
*France*

RESUME

L'objectif de ce travail est de proposer un modèle réaliste et opérationnel pour mesurer le « risque de dérive » associé à la construction de tables de mortalité prospectives. Une application du modèle à l'évaluation de l'engagement d'un engagement de retraite est proposée. Le modèle présenté est construit sur la base d'un modèle de Lee-Carter. Les tables prospectives stochastiques sont obtenues en modélisant l'incertitude attachée au paramètre tendanciel du modèle.

ABSTRACT

The aim of this paper is to propose a realistic and operational model to quantify the systematic risk of mortality included in an engagement of retirement. The model presented is built on the basis of model of Lee-Carter. The stochastic prospective tables thus built make it possible to project the evolution of the random mortality rates in the future and to quantify the systematic risk of mortality.

* fplanchet@winter-associes.fr

**SOMMAIRE**

# 1. INTRODUCTION

Les analyses prospectives de mortalité conduisent à anticiper les évolutions futures des taux de décès aux différents âges. Dans les modèles maintenant classiques de construction de tables prospectives, comme le modèle de Lee-Carter (voir notamment LEE et CARTER [1992], LEE [2000], SITHOLE et al. [2000]) ou les modèles poissoniens (*cf.* BROUHNS et al. [2002] et PLANCHET et THEROND [2006] pour une présentation et une discussion de ces modèles), la dérive de mortalité future est anticipée à partir des observations passées.

Même en admettant qu'il est légitime de prolonger dans les années à venir les tendances observées par le passé (on pourra se reporter à CAREY et TULAPURKAR [2003] pour des analyses intégrant des considérations biologiques et environementales, ainsi que GUTTERMAN et VANDERHOOF [1999] pour une discussion sur ce point), plusieurs sources d'incertitude viennent perturber la détermination de la tendance future : choix de la période d'observation, fluctuations stochastique des taux de mortalité, événements exceptionnels, *etc.*. Cette incertitude fait peser sur les assureurs de rentes viagères et les régimes de retraite un risque systématique (non mutualisable) dont l'impact financier peut être très important.

Ainsi, en France, la récente actualisation des tables utilisées par les assureurs pour le provisionnement des rentes viagères illustre les difficultés d'une telle anticipation et les enjeux financiers associés : par rapport aux tables TPG[1] 1993 en vigueur jusqu'au 31/12/2006, les nouvelles tables TGH 05 et TGF 05 qui entrent en vigueur le 01/01/2007 conduisent à des majorations de provision parfois supérieures à 20%, comme l'illustre le tableau suivant :

| Age | Génération | TPG 1993 | Femmes | Femmes / TPG | Hommes | Hommes / TPG |
|---|---|---|---|---|---|---|
| 50 | 1955 | 26,81647 | 28,40552 | 5,9% | 26,75507 | -0,2% |
| 55 | 1950 | 24,26368 | 25,95575 | 7,0% | 24,07474 | -0,8% |
| 60 | 1945 | 21,50832 | 23,30185 | 8,3% | 21,25828 | -1,2% |
| 65 | 1940 | 18,53412 | 20,39677 | 10,0% | 18,22126 | -1,7% |
| 70 | 1935 | 15,39467 | 17,28922 | 12,3% | 15,08772 | -2,0% |
| 75 | 1930 | 12,25679 | 14,08680 | 14,9% | 12,05698 | -1,6% |
| 80 | 1925 | 9,35194 | 10,96271 | 17,2% | 9,12890 | -2,4% |
| 85 | 1920 | 6,88306 | 8,15548 | 18,5% | 6,64827 | -3,4% |
| 90 | 1915 | 4,93310 | 5,89309 | 19,5% | 4,73880 | -3,9% |
| 95 | 1910 | 3,46780 | 4,29408 | 23,8% | 3,40109 | -1,9% |

Fig. 1 : *Comparaison des coefficients de provisionnement TPG 1993 et TGH/TGF 05*

Dans ce contexte il apparaît opportun de rechercher à mesurer le risque associé à cette erreur

[1] Tables obtenues sur la base de la mortalité de la population féminine sur la période 1961-1987, utilisées depuis le 1er juillet 1993.

d'anticipation et de quantifier son impact en terme de provisions pour un régime de rentiers.

On utilise pour cela dans la présente étude le modèle de Lee-Carter (voir notamment LEE et CARTER [1992], LEE [2000], SITHOLE et al. [2000]) pour construire une surface de mortalité $\mu(x,t)$. Après un ajustement des taux passés, les taux de mortalité pour les années futures se déduisent classiquement de l'extrapolation de la composante temporelle. On peut noter que l'utilisation de la variante log-Poisson (*cf.* BROUHNS et al. [2002]) conduirait à des résultats très proches, qui ne seront pas repris ici.

A partir de ce modèle de référence, on propose un modèle stochastique de mortalité en considérant que le taux de mortalité futur $\mu(x,t)$ est lui-même aléatoire, et que donc $\mu(x,t)$ est un processus stochastique (comme fonction de $t$ à $x$ fixé). L'aléa est introduit de manière à capturer l'incertitude sur l'estimation de la tendance future de la composante temporelle des taux de mortalité.

Après avoir construit un jeu de tables prospectives sur des données nationales à l'aide de ce modèle, nous l'utilisons pour calibrer l'incertitude sur la dérive anticipée et appliquons le modèle ainsi obtenu pour déterminer la distribution de l'engagement d'un régime de rentes. Les conséquences en terme d'anamyse du risque pesant sur le régime et de provisionnement sont examinées.

Le présent article complète l'analyse présentée dans PLANCHET et al. [2006], auquel le lecteur pourra se référer pour les détails méthodologiques des modèles de base.

## 2. LE MODELE DE MORTALITE

### 2.1. PRESENTATION

Le modèle retenu pour constuire les tables prospectives est un modèle stochastique adapté du modèle de Lee-Carter (LEE et CARTER [1992]). On rappelle que la modélisation proposée pour le taux instantané de mortalité dans Lee-Carter est la suivante :

$$\ln \mu_{xt} = \alpha_x + \beta_x k_t + \varepsilon_{xt}, \qquad (1)$$

en supposant les variables aléatoires $\varepsilon_{xt}$ indépendantes, identiquement distribuées selon une loi $N(0,\sigma^2)$ et que l'on dispose d'un historique $t_m \leq t \leq t_M$. La question de l'ajustement des paramètres du modèle n'est pas abordée ici. Le lecteur intéressé pourra se reporter aux nombreuses références sur le sujet (citées par exemple dans PLANCHET et THEROND [2006]) .

Une fois ajustée la surface de mortalité sur les données passées, il reste à modéliser la série $(k_t)$ pour extrapoler les taux futurs ; pour cela, on utilise la modélisation la plus simple que l'on puisse imaginer, une régression linéaire en supposant une tendance affine :

$$k_t^* = at + b + \gamma_t, \tag{2}$$

avec $(\gamma_t)$ un bruit blanc gaussien de variance $\sigma_\gamma$. On obtient ainsi des estimateurs $\hat{a}$ et $\hat{b}$ qui permettent de construire des surfaces projetées en utilisant simplement $k_t^* = \hat{a}t + \hat{b}$.

Afin de simplifier l'écriture des formules à venir, on pose $\tau = t - t_m + 1$ et $T = t_M - t_m + 1$, ce qui conduit aux expressions :

$$\hat{a} = \frac{\frac{1}{T}\sum \tau k_\tau - \frac{T+1}{2}\bar{k}}{T^2 - 1} \text{ et } \hat{b} = \bar{k} - \frac{T+1}{2}\hat{a}, \tag{3}$$

avec $\bar{k} = \frac{1}{T}\sum k_\tau = \frac{1}{T}\sum k_t$. De plus, le vecteur $(\hat{a}, \hat{b})$ est distribué selon une loi normale d'espérance $(a, b)$ et de variance :

$$\Sigma = \frac{12\sigma_\gamma^2}{T\left(T^2-1\right)}\begin{bmatrix} 1 & -\frac{T+1}{2} \\ -\frac{T+1}{2} & \frac{(T+1)(2T+1)}{6} \end{bmatrix} \tag{4}$$

On peut donc construire des réalisations de la mortalité future en effectuant des tirages dans la loi du vecteur $(\hat{a}, \hat{b})$. La variable $k_t^*$ ainsi obtenue est telle que $E\left(k_t^*\right) = k_t$. On obtient alors des réalisations des taux instantanés de sortie *via* :

$$\mu_{xt}^* = \mathbf{exp}\left(\alpha_x + \beta_x k_t^*\right). \tag{5}$$

Comme $k_t^* = \hat{a}t + \hat{b}$ est une variable gaussienne d'espérance $k_t = at + b$ et de variance :

$$\sigma_\tau^2 = \frac{12\sigma_\gamma^2}{T\left(T^2-1\right)}\left(\tau^2 - \tau(T+1) + \frac{(T+1)(2T+1)}{2}\right), \tag{6}$$

on note que :

$$E\left(\mu_{xt}^{*}\right)=E\left(\mathbf{exp}\left(\alpha_x+\beta_x k_t^{*}\right)\right)=\mathbf{exp}\left(\alpha_x+\beta_x k_t+\frac{\beta_x^2\sigma_t^2}{2}\right), \tag{7}$$

et donc :

$$E\left(\mu_{xt}^{*}\right)=\mu_{xt}\,\mathbf{exp}\left(\frac{\beta_x^2\sigma_t^2}{2}\right)>\mu_{xt}. \tag{8}$$

Le modèle stochastique a donc tendance à surestimer les taux de sortie par rapport à la surface de référence fournie par le modèle de Lee-Carter. Compte tenu du notre objectif de « perturber » la surface de mortalité, mais sous l'hypothèse que celle-ci définie correctement la tendance future espérée des taux instantanés de décès cette propriété du modèle est pénalisante et il convient d'adapter l'approche proposée.

PLANCHET et al. [2006] proposent une version corrigée du biais du modèle définie par :

$$\mu_{xt}^{*}=\mathbf{exp}\left(\alpha_x-\frac{\beta_x^2\sigma_t^2}{2}+\beta_x k_t^{*}\right) \tag{9}$$

Cette version du modèle satisfait par construction $E\left(\mu_{xt}^{*}\right)=\mu_{xt}$. Elle apparaît donc cohérente avec l'objectif recherché. Toutefois, on peut lui reprocher de déformer de manière arbitraire la distribution des taux stochastiques ; en effet, si le modèle $k_t^{*}=\hat{a}t+\hat{b}$ est pertinent, alors les taux de sortie $\mu_{xt}^{*}$ réellement observés seront bien issus du modèle $\mu_{xt}^{*}=\mathbf{exp}\left(\alpha_x+\beta_x k_t^{*}\right)$ et non de la version corrigée du biais.

On utilise ici une approche différente et *a priori* plus naturelle consistant à utiliser comme surface de référence déterministe la surface moyenne du modèle stochastique, soit $E\left(\mu_{xt}^{*}\right)=\mathbf{exp}\left(\alpha_x+\beta_x k_t+\frac{\beta_x^2\sigma_t^2}{2}\right)$. En effet, si le mécanisme aléatoire dont sont issus les taux de décès est bien associé à $k_t^{*}=\hat{a}t+\hat{b}$, alors la mortalité de référence déterministe est bien définie par la surface moyenne ci-dessus, et non plus par la surface de Lee-Carter.

## 2.2. APPLICATION NUMERIQUE

On présente ici les résultats obtenus tout d'abord sur la famille de tables prospectives proposée puis, dans un second temps, les conséquences en terme de valorisation de l'engagement du régime de rentes.

La table prospective utilisée dans cette étude est celle utilisée dans PLANCHET et al. [2006] ; elle est construite à partir des tables du moment fournies par l'INED[2] dans MESLE et VALLIN [2002] et conduit à la surface de mortalité Lee-Carter suivante :

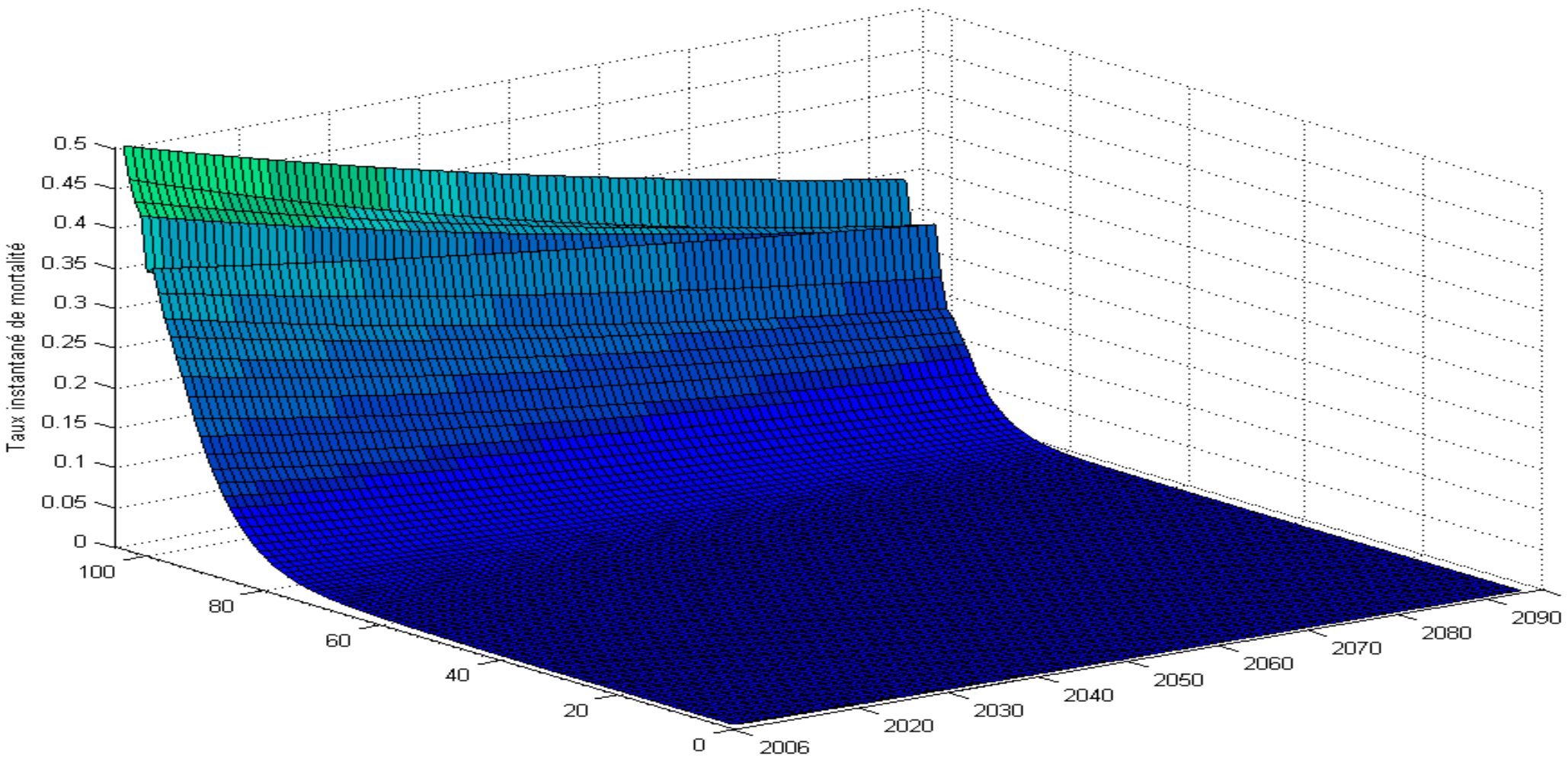

Fig. 2 : *Surface de mortalité ajustée par Lee-Carter*

La surface de référence ajustée de notre modèle stochastique est présentée sur le graphe ci-dessous :

[2] Ces tables sont disponibles sur http://www.ined.fr/publications/cdrom_vallin_mesle/Tables-de-mortalite/Tables-du-moment/Tables-du-moment-XX.htm

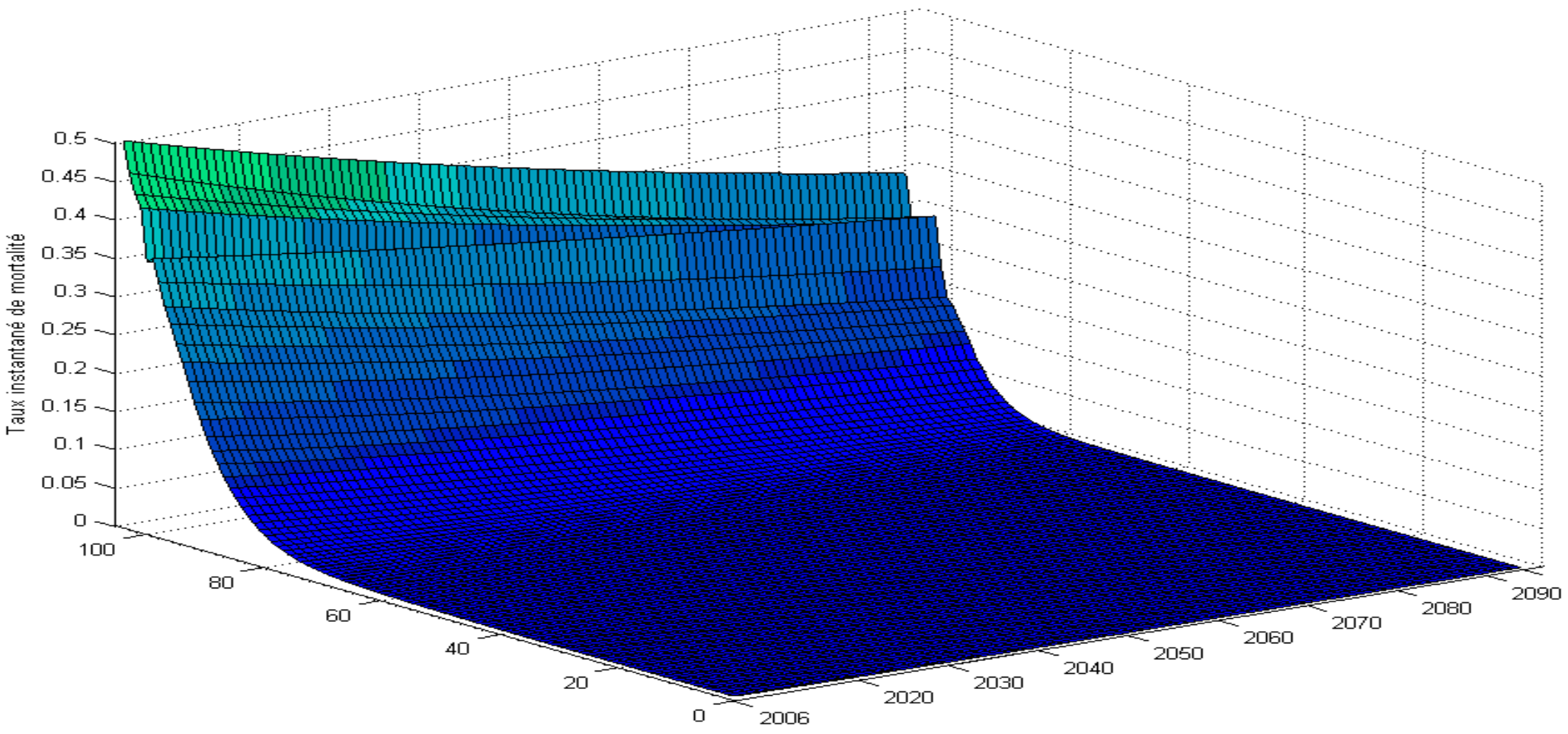

Fig. 3 : *Surface de mortalité de référence du modèle stochastique*

On remarque que la différence entre les deux graphes est très faible car du fait de la faible valeur des coefficients $\beta_x$, $\mathbf{exp}\left(\frac{\beta_x \sigma_t^2}{2}\right)$ est proche de 1. En ce qui concerne le volet prédictif du modèle, nous obtenons sur nos données :

$$\hat{b} = 49.38604,\ \hat{a} = -2.05775 \text{ et } \hat{\sigma}_\gamma = 3.98227882$$

$$\hat{\sigma}_b = 1.18058,\ \hat{\sigma}_a = 0.04282 \text{ et } \hat{\sigma}_t^{\,2} = 1.39367413 + 0.00183378t^2 - 0.08802152t$$

Une centaine de tirages de trajectoires de $k_t$ permet d'obtenir le graphe suivant :

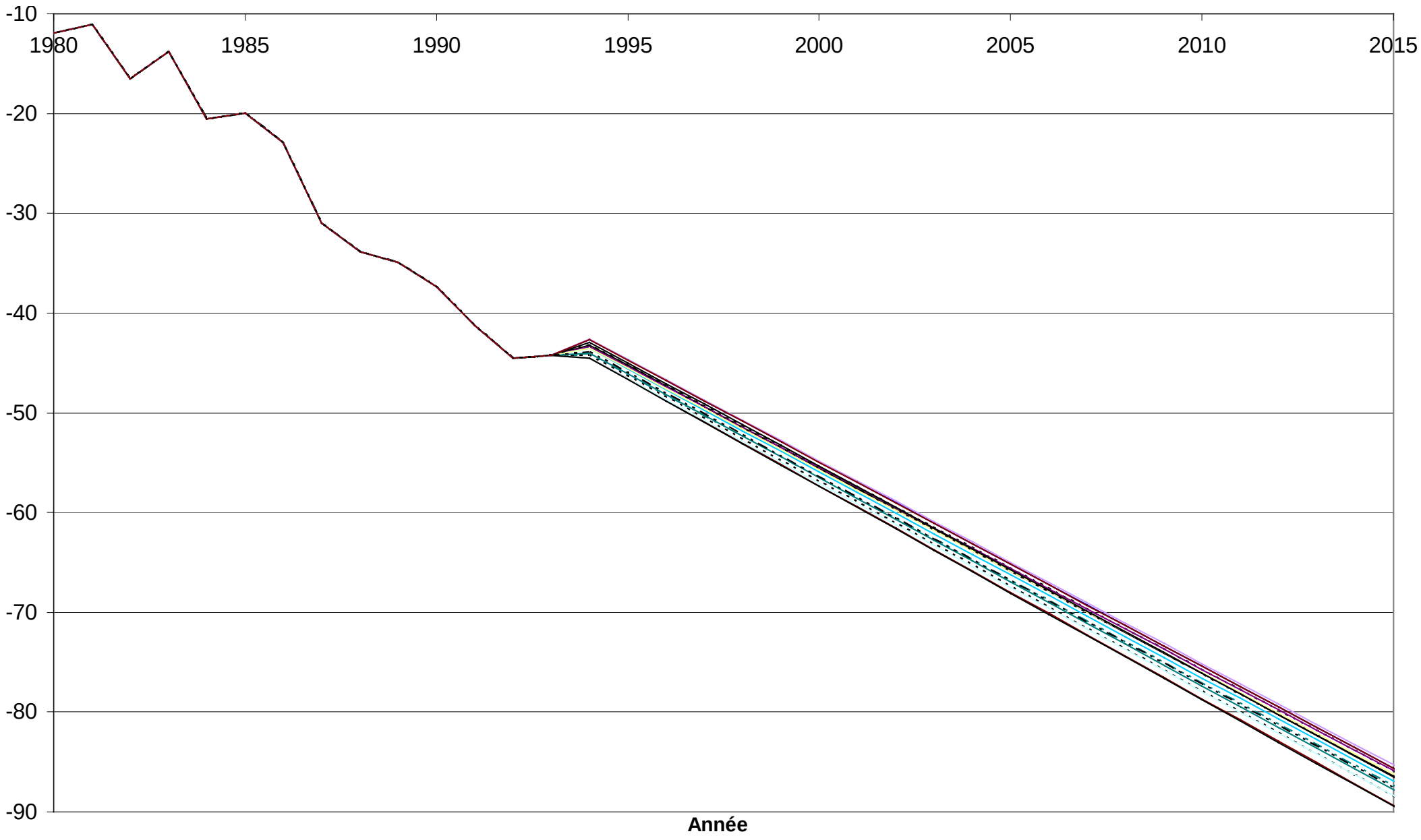

Fig. 4 : *Simulation de trajectoires de la tendance*

On constate sur le graphe que les différentes trajectoires simulées de $k_t$ prennent quasiment la même valeur en 2006. Ceci montre que le modèle tire sa volatilité de la dérive de la mortalité et non d'un saut brutal de mortalité entre 2005 et 2006.

Si l'on cherche à comparer la volatilité de la dérive temporelle modélisée dans cet article avec celle du modèle utilisé dans PLANCHET et al. [2006], on obtient le graphe suivant :

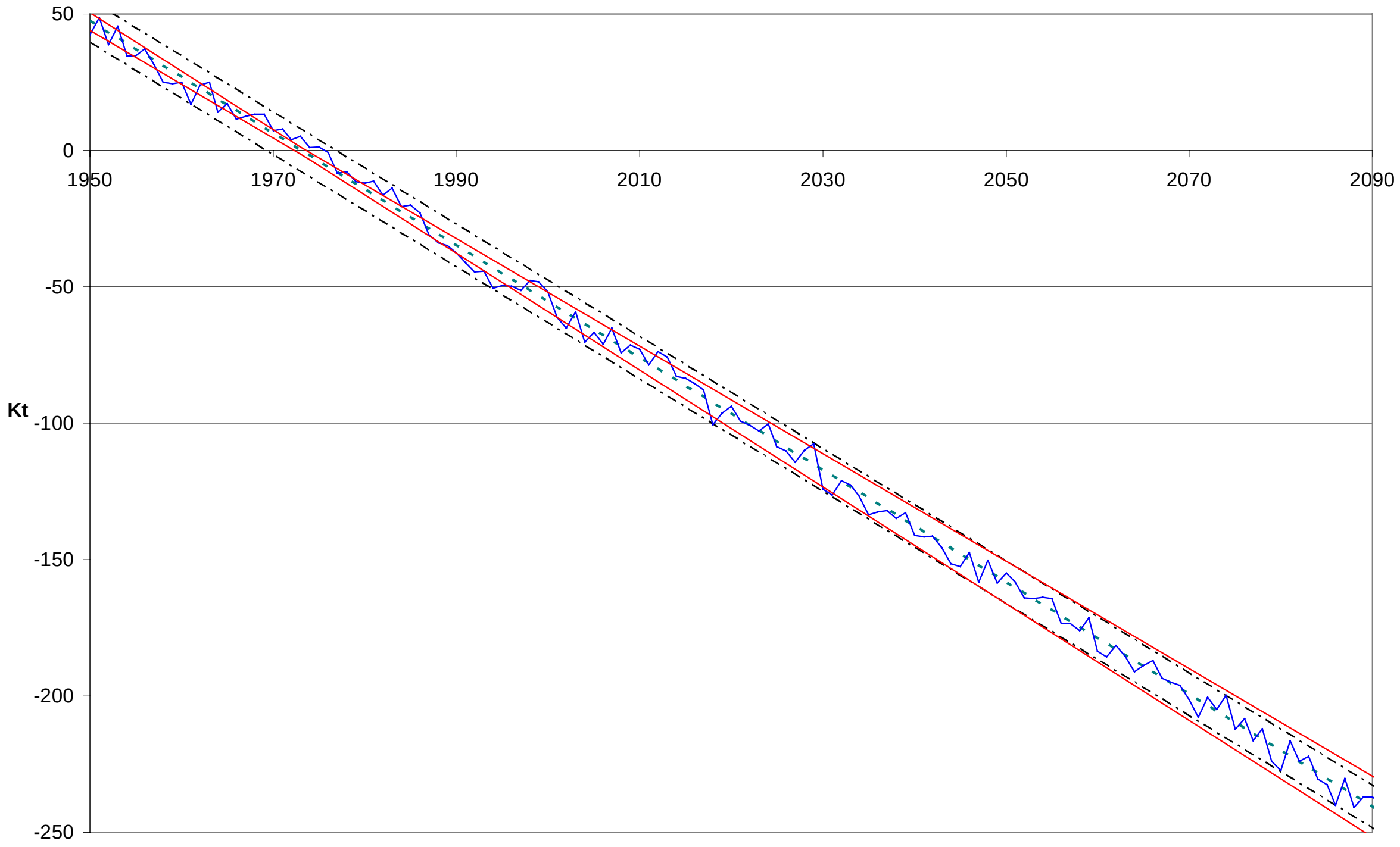

Fig. 5 : *« couloir de variation » de la tendance*

Alors que dans PLANCHET et *al.* [2006] le modèle utilisé ne prenait en compte que l'oscillation de la dérive temporelle $k_t$ autour de sa moyenne, le modèle actuel simule l'erreur d'ajustement qui a pu être faite sur la dérive temporelle passée.

Si dans PLANCHET et *al.* [2006] les différentes trajectoires de $k_t$ étaient situées entre deux droites parallèles délimitant le couloir de variation de la tendance à 95 %, on constate sur le graphique ci-dessus que les trajectoires de la dérive temporelle sont situées entre deux droites s'éloignant l'une de l'autre avec le temps. Les variations de la tendance temporelle et celles du taux instantané de mortalité étant comparables, le graphique ci-dessus traduit l'hypothèse du modèle : la mortalité stochastique tire sa volatilité non pas des fluctuations annuelles qui, comme nous l'avons démontré dans PLANCHET et *al.* [2006] restent faibles, mais plutôt de la dérive aléatoire de la mortalité qui n'est vraiment observable que dans le futur.

## 3. APPLICATION A UN REGIME DE RENTES VIAGERES

### 3.1. PROBLEMATIQUE

Dans la suite, nous utiliserons pour les applications numériques un portefeuille constitué de 374 rentiers de sexe féminin âgés en moyenne de 63,8 ans au 31/12/2005. La rente annuelle moyenne s'élève à 5,5 k€. Avec un taux d'escompte des provisions de 2,5 %, la provision mathématique initiale, s'élève à 37,9 M€ avec la table prospective déterminée *supra*.

Nous noterons dans la suite de l'article :

- ✓ $L_0$ le montant des provisions mathématiques à la date initiale,
- ✓ $\tilde{F}_t$ le flux de prestation (aléatoire) à payer à la date *t*,
- ✓ *i* le taux (discret) d'escompte des provisions mathématiques,
- ✓ **J** l'ensemble des individus,
- ✓ *x*(*j*) l'âge en 0 de l'individu *j* et $r_j$ le montant de sa rente annuelle.

La problématique est décrite de manière détaillée dans PLANCHET et *al.* [2006] ; nous retiendrons ici que l'on s'intéresse à la loi de du montant de l'engagement du régime :

$$\Lambda = \sum_{t=1}^{\infty} \tilde{F}_t (1+i)^{-t} = \sum_{t=1}^{\infty} \frac{1}{(1+i)^t} \sum_{j \in \boldsymbol{J}} r_j * \mathbf{1}_{]t;\infty[}\left(T_{x(j)}\right) \tag{10}$$

qui est une variable aléatoire telle que $E(\Lambda) = L_0$ avec $L_0 = \sum_{t=1}^{\infty} F_t (1+i)^{-t}$.

Lorsque la mortalité future est connue (déterministe), l'analyse de la loi de $\Lambda$ revient à mesurer les fluctuations d'échantillonnage qui est associé au « risque de volatilité » tel que décrit dans la modélisation proposées par les QIS 2 et 3 par exemple (*cf.* CEIOPS [2006]). Dans le contexte d'une mortalité stochastique, cette analyse nous fournit un moyen de quantifier le risque systématique qui vient s'ajouter à ce risque de base : ici nous pourrons donc mesurer ainsi le risque associé au choix de la dérive.

La méthode retenue consiste à simuler les durées de survie des rentiers, $T_{x(j)}, j \in J$, à calculer des réalisations $\lambda_1, ..., \lambda_n$ de $\Lambda$ puis à déterminer la fonction de répartition empirique de l'engagement.

La provision $L_0$ est approchée par $\overline{\lambda} = \frac{1}{N}\sum_{n=1}^{N}\lambda_n$. La variance de l'engagement est estimée par $\frac{1}{N-1}\sum_{n=1}^{N}(\lambda_n - L_0)^2$ ; on calcule également le coefficient de variation empirique :

$$cv = \frac{\sqrt{\frac{1}{N-1}\sum_{n=1}^{N}(\lambda_n - L_0)^2}}{\frac{1}{N}\sum_{n=1}^{N}\lambda_n}, \qquad (11)$$

qui fournit un indicateur de la dispersion de l'engagement et dans une certaine mesure de sa « dangerosité ».

## 3.2. RESULTATS

La distribution empirique de l'engagement, représentée ici avec la distribution de référence dans le cas déterministe (avec 20 000 tirages) est présentée ci-dessous :

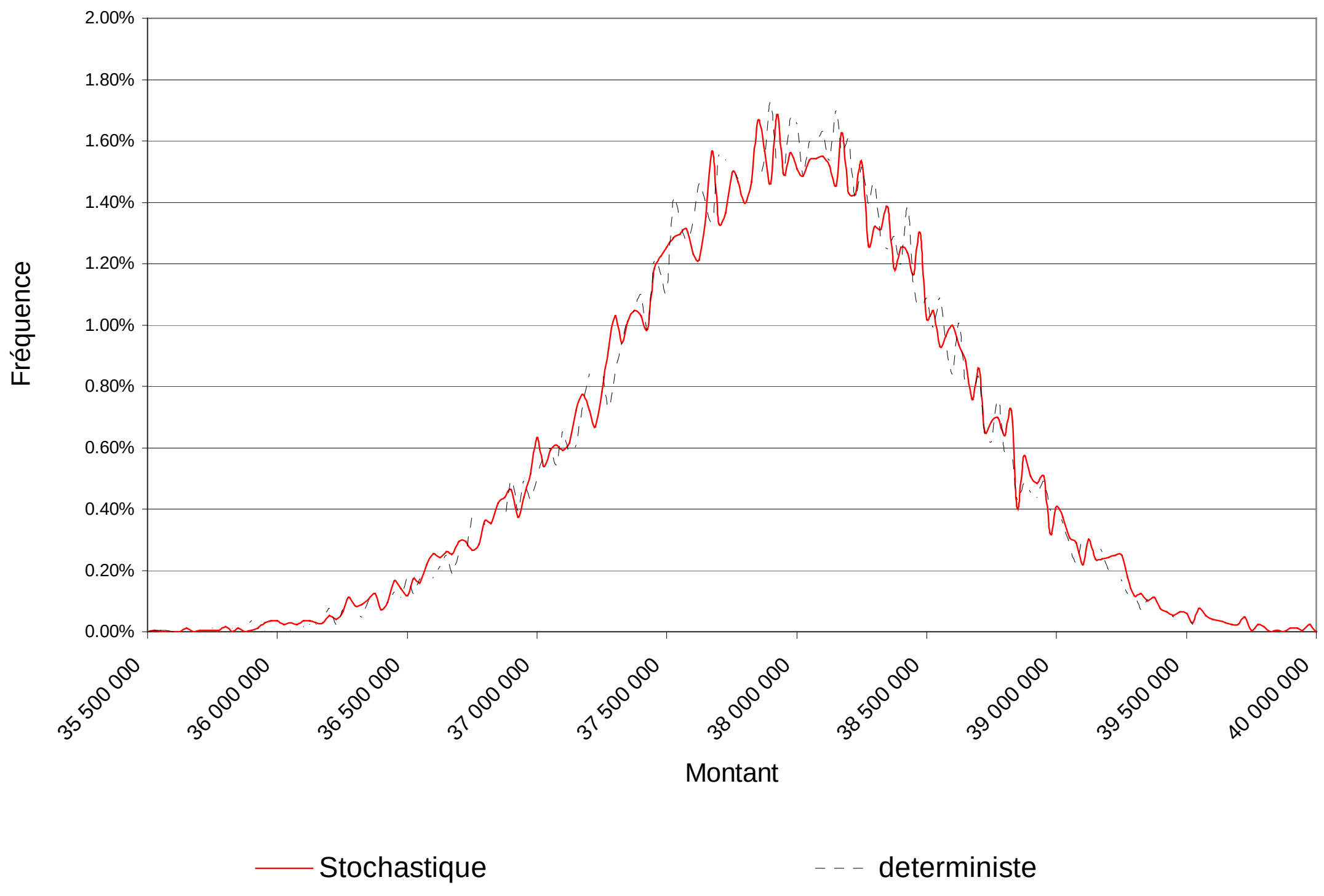

Fig. 6 : *Distribution empirique de l'engagement*

Les résultats détaillés sont repris ci-après :

| | Déterministe surface Lee Carter | Déterministe | Stochastique |
|---|---|---|---|
| Espérance | 37 937 707 | 37 937 707 | 37 937 720 |
| Ecart-type | 626 918 | 626 918 | 645 601 |
| Borne inférieure de l'intervalle de confiance | 36 625 000 | 36 625 000 | 36 600 000 |
| Borne supérieure de l'intervalle de confiance | 39 075 000 | 39 075 000 | 39 150 000 |
| Coefficient de variation | 1,65 % | 1,65 % | 1,70 % |

On remarque tout d'abord que la modification de la surface de référence utilisée dans PLANCHET et *al.* [2006] n'a pas d'impact. Ceci est du au fait déjà observé que pour tout $x$ et pour tout $t$ $\mathbf{exp}\left(\frac{\beta_x\sigma_t^2}{2}\right)$ est proche de 1.

Sur un portefeuille de petite taille, l'impact de la mortalité stochastique sur l'engagement n'est pas très important. En effet le coefficient de variation de l'engagement stochastique n'est que 2,9 % plus élevé que celui de l'engagement déterministe.

Cependant comme nous l'avions montré dans PLANCHET et *al.* [2006] la taille du portefeuille est un paramètre important à prendre en compte. En effet si le niveau absolu de risque systématique ne dépend pas de la taille du portefeuille, il n'en va pas de même pour le risque mutualisable. La part de variance expliquée par la composante stochastique de la mortalité augmente donc avec la taille du portefeuille ; afin de mesurer cet effet, on construit un portefeuille fictif en répliquant le portefeuille de base $n$ fois. En observant que l'on obtient ainsi une décomposition de l'engagement total $\Lambda$ en la somme de $n$ variables i.i.d. $\Lambda^{(1)},..,\Lambda^{(n)}$ on trouve que :

$$V\left[E\left(\Lambda|\Pi\right)\right]=n^2V\left[E\left(\Lambda^{(1)}|\Pi\right)\right] \text{ et } E\left[V\left(\Lambda|\Pi\right)\right]=nE\left[V\left(\Lambda^{(1)}|\Pi\right)\right], \quad (12)$$

ce qui conduit, avec des notations évidentes, à :

$$\omega_n=\left(1+\frac{1}{n}\left(\frac{1}{\omega}-1\right)\right)^{-1}, \quad (13)$$

avec $\omega=\frac{\boldsymbol{V}\left[\boldsymbol{E}\left(\Lambda|\Pi\right)\right]}{\boldsymbol{V}\left[\Lambda\right]}$. Si l'on multiplie le portefeuille par 30 on obtient les résultats suivants :

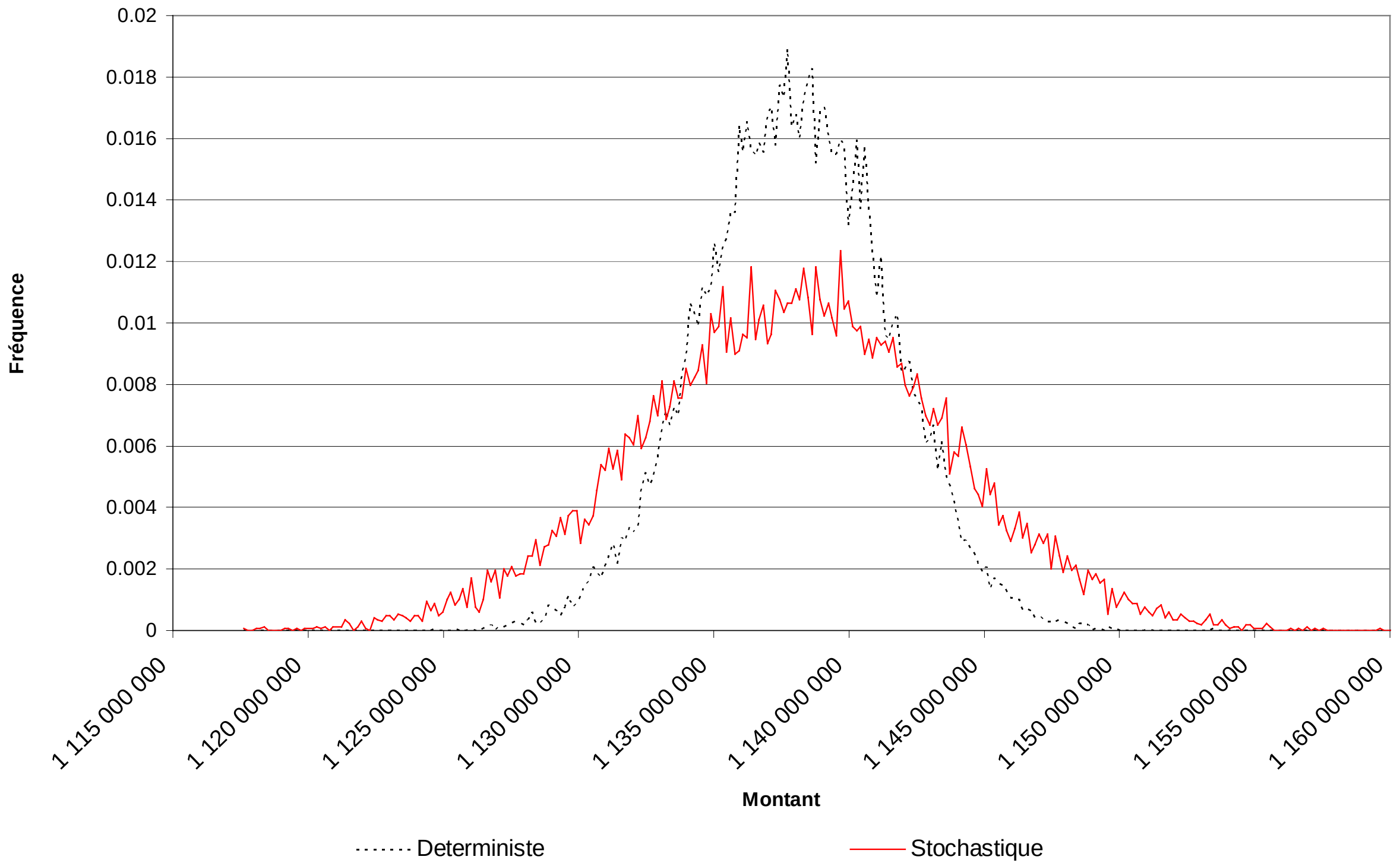

Fig. 7 : *Distribution empirique de l'engagement (taille x30)*

Les résultats détaillés sont repris ci-après :

| | **Déterministe** | **Stochastique** |
|---|---|---|
| Espérance | 1 138 008 113 | 1 138 076 960 |
| Ecart-type | 5 592 212 | 3 410 560 |
| Borne inférieure de l'intervalle de confiance | 1 131 130 658 | 1 126 780 658 |
| Borne supérieure de l'intervalle de confiance | 1 144 480 658 | 1 148 830 658 |
| Coefficient de variation | 0,30 % | 0,49 % |

La prise en compte de la mortalité stochastique augmente le coefficient de variation de 63 %.

Dans une approche « valeur à risque » (*VaR*), on trouve que le quantile à 75 % de la distribution de l'engagement est de 1 142 M€ dans le cas stochastique, ce qui est supérieur à la valeur obtenue dans le cas déterministe, soit 1 140 M€. En d'autres termes, la prise en compte du risque de dérive conduit ici (en suivant une approche *VaR* pour le calcul de la provision) à augmenter le montant provisionné de 0,17 %.

La prise en compte du risque de dérive a également pour conséquence de doubler l'imprécision dans l'évaluation de l'engagement[3] qui, au niveau de confiance de 95 %, passe de 0,7 % environ à près de 1,3 %

[3] L'imprécision est mesurée par la demi-longueur relative de l'intervalle de confiance à 95 %.

Le graphique du « couloir de variation » de la tendance ayant mis en évidence le faible impact de la dérive de la mortalité dans le futur proche, le coefficient d'actualisation va jouer un rôle important dans la volatilité de la distribution de l'engagement.

En effet l'engagement stochastique tire sa volatilité des rentes qui vont être servies pendant longtemps. Or plus la rente à une durée de vie importante, plus le coefficient d'actualisation qui s'y applique est important. En fixant le taux d'actualisation à 0 on obtient les résultats suivant :

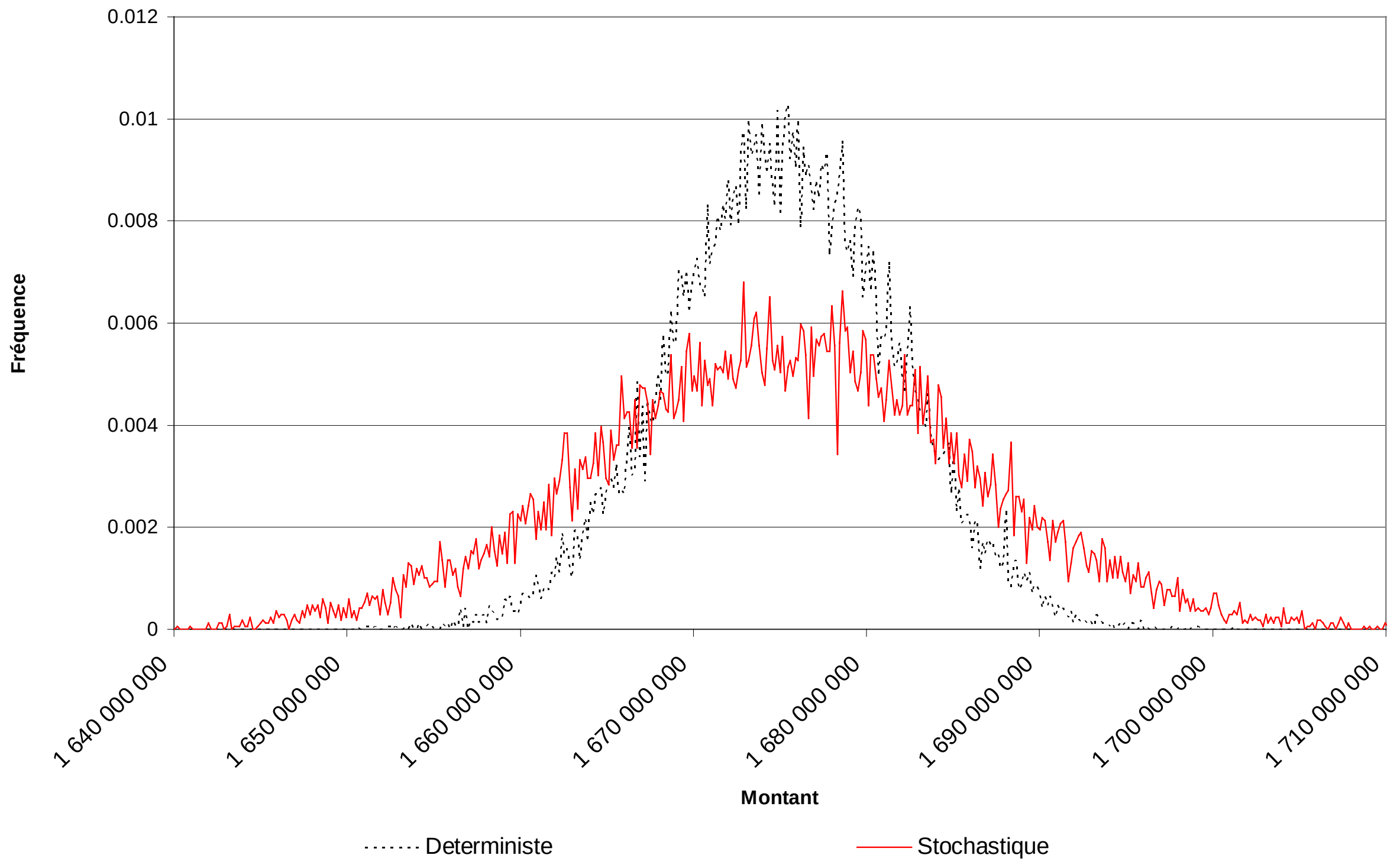

Fig. 8 : *Distribution empirique de l'engagement (taux technique nul)*

Les résultats détaillés sont repris ci-après :

| | Déterministe | Stochastique |
|---|---|---|
| Espérance | 1 675 256 646 | 1 675 129 393 |
| Ecart-type | 6 389 153 | 10 893 496 |
| Borne inférieure de l'intervalle de confiance | 1 662 554 600 | 1 653 554 600 |
| Borne supérieure de l'intervalle de confiance | 1 687 754 600 | 1 696 454 600 |
| Coefficient de variation | 0,38 % | 0,65 % |

En prenant un taux technique nul on augmente le coefficient de variation de l'engagement déterministe de 27 % et le coefficient de variation de l'engagement stochastique de 33 %. Ainsi la prise en compte de la mortalité stochastique induit une plus grande sensibilité de l'engagement au taux d'actualisation.

## 4. CONCLUSION

Dans PLANCHET et al. [2006] nous avions mis en évidence le fait que l'évaluation de l'engagement d'un régime de rentiers, l'aléa associé aux fluctuations des taux de décès futurs autour de leur tendance n'avait qu'un faible impact que l'on pouvait légitimement négliger.

Le modèle mis en œuvre ici permet de montrer que, si ces régimes sont particulièrement sensibles au risque d'erreur de spécification de la tendance future, l'incertitude sur ce point incluse dans les données passées ne suffit pas à expliquer les évolutions récentes observées.

Incidemment, on insistera sur le fait que l'on a fait ici le choix de modéliser l'aléa sur la tendance future de mortalité à partir des seules fluctuations du paramètre temporel $k_t$ ; cette approche est justifiée par le fait que l'on cherche à identifier la volatilité des taux de décès non associée aux fluctuations d'échantillonnage. Compte tenu des effectifs très importants du groupe utilisé pour construire les tables utilisées ici (la population française), on peut en effet considérer qu'il n'y a plus de fluctuations d'échantillonnage significatives dans l'estimation des taux bruts[4] et que donc il est légitime de considérer la surface passée fixe.

Une approche alternative consisterait à prendre en compte globalement l'incertitude sur le paramètre vectoriel $\theta = (\alpha, \beta, k)$ pour générer des surfaces de mortalité. Cette démarche est techniquement simple à mettre en œuvre à la condition d'utiliser la variante log-Poisson du modèle de Lee-Carter, dans laquelle l'estimation de $\theta$ est effectuée par maximum de vraisemblance, puisque, alors, on peut déterminer l'information de Fischer associée et utiliser la normalité asymptotique de l'estimateur[5].

Toutefois, elle nous semble moins pertinente dans le cas présent, notre objectif étant d'isoler le risque associé à une incertitude structurelle sur les taux et non à une incertitude d'estimation.

En conclusion, on peut retenir que le régime de rente est par contre soumis à un risque de modèle important. Il est ainsi possible de reformuler les conséquences du passage des tables TPG 1993 aux tables[6] TGH 05 évoquées en introduction en rapprochant l'évolution anticipée de l'espérance de vie à 60 ans dans les deux modèles prospectifs. On obtient ainsi :

---

[4] Sur les adaptations du modèle de Lee-Carter au cas de petits échantillons pour lesquels cette hypothèse n'est plus vérifiée, on pourra se reporter à PLANCHET et LELIEUR [2006].
[5] Voir par exemple HAAS [2006] pour une utilisation de cette approche dans le cas de swaps de mortalité.
[6] On rappelle que les TPG 1993 sont des tables féminines.

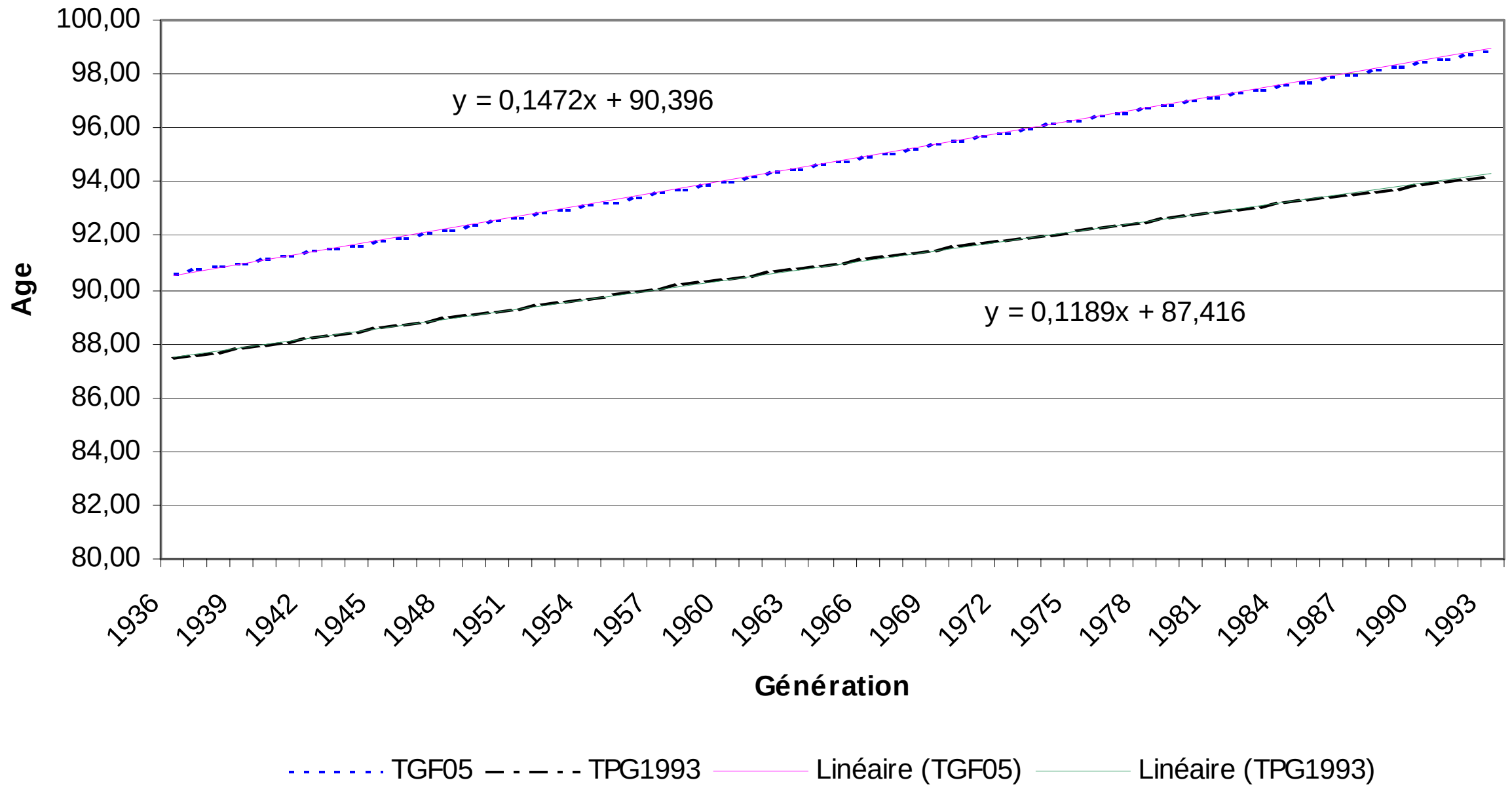

Fig. 9 : *Evolution anticipée de l'espérance de vie à 60 ans*

On observe que non seulement les niveaux absolus diffèrent sensiblement, mais également que la vitesse de croissance de l'espérance de vie à 60 ans a été sensiblement sous-estimée en 1993 : alors que les TPG 1993 anticipent une augmentation de 1,4 mois par an, les tables TGF 05 prévoient une dérive de 1,8 mois par an, soit 23 % de plus.

Ceci illustre la difficulté à anticiper la tendance de dérive de la mortalité future à partir de données historiques[7]. Dans ce contexte, on peut imaginer d'utiliser comme paramètre de contrôle du modèle, en le fixant comme une contrainte *ex ante* par exemple l'espérance de vie à un âge donné (60 ans) et son évolution future. Cette approche, développée dans PLANCHET [2007], permet de quantifier par exemple l'impact sur les charges du régime d'une erreur de 0,1 mois/an sur la vitesse de dérive de cette espérance et d'intégrer explicitement des indicateurs de l'impact sur l'évaluation de l'engagement du régime d'une erreur de modèle. Le modèle proposé dans PLANCHET [2007] permet ainsi de proposer une valorisation de la charge associée au risque de longévité à environ 6 % de la provision mathématique des rentes. Cet ordre de grandeur est à rapprocher du taux de 0,17 % que l'on déduit de l'incertitude associée au caractère stochastique « endogène » de la mortalité dans le modèle de Lee-Carter ou ses variantes.

On peut enfin rappeler que ce type d'approche fournit un cadre opérationnel pour répondre aux exigences des futures dispositions Solvabilité 2 (et également dans le contexte de l'IFRS 4 phase 2, quoi que sur ce point les normes comptables soient moins exigeantes).

---

[7] Les populations de référence utilisées pour les 2 séries de tables diffèrent, mais on obtiendrait les même conclusions en utilisant des tables prospectives INSEE à la place des TGF 05.

## BIBLIOGRAPHIE